**Optimizing normal tissue sparing via spatiotemporal optimization under equivalent tumor-radical efficacy**

Nimita Shinde, Wangyao Li, Ronald C Chen, and Hao Gao*

Department of Radiation Oncology, University of Kansas Medical Center, USA

* Corresponding author:

Hao Gao, Department of Radiation Oncology, University of Kansas Medical Center, USA.

Email address: hgao2@kumc.edu

**Conflict of Interest Statement**

None.

**Ethical Statement:** This research was carried out under Human Subject Assurance Number 00003411 for University of Kansas in accordance with the principles embodied in the Declaration of Helsinki and in accordance with local statutory requirements. Consent was given for publication by the participants of this study.

**Acknowledgment**

The authors are very thankful for the valuable comments from reviewers. This research is partially supported by the NIH grants No. R37CA250921, R01CA261964, and a KUCC physicist-scientist recruiting grant.

**Abstract.**

**Objective:** Spatiotemporal optimization in radiation therapy involves determining the optimal number of dose delivery fractions (temporal) and the optimal dose per fraction (spatial). Traditional approaches focus on maximizing the biologically effective dose (BED) to the target while constraining BED to organs-at-risk (OAR), which may lead to insufficient BED for complete tumor cell kill. This work proposes a formulation that ensures adequate BED delivery to the target while minimizing BED to the OAR.

**Approach:** A spatiotemporal optimization model is developed that incorporates an inequality constraint to guarantee sufficient BED for tumor cell kill while minimizing BED to the OAR. The model accounts for tumor proliferation dynamics, including lag time (delay before proliferation begins) and doubling time (time for tumor volume to double), to optimize dose fractionation.

**Main results:** The proposed formulation is implemented for proton modality. The performance of our method is evaluated for varying lag times and doubling times. The results show that the mean BED to the target consistently meets the minimum requirement for tumor cell kill. Additionally, the mean BED to the OAR varies based on tumor proliferation dynamics. In the prostate case with lag time of 7 days and doubling time of 2 days, it is observed that the mean BED delivered to femoral head is lowest at approximately 20 fractions, making this an optimal choice. While in the head-and-neck case, the mean BED to the OAR decreases as the number of fractions increases, suggesting that a higher number of fractions is optimal. Thus, the proposed model effectively determines the optimal fractionation strategy under different tumor proliferation conditions.

**Significance:** A spatiotemporal optimization model is presented that minimizes BED to the OAR while ensuring sufficient BED for tumor cell kill. By incorporating tumor lag time and doubling time, the approach identifies optimal number of fractions. This model can be extended to support hyperfractionation or accelerated fractionation strategies, offering a versatile tool for clinical treatment planning.

## 1. Introduction

Radiation therapy [1-3] aims to maximize tumor damage while minimizing damage to surrounding normal tissues and organs-at-risk (OAR). This balance is achieved through temporal dose distribution (fractionating the treatment over multiple sessions) [4] and spatial dose modulation (optimizing dose delivery in each session) [5].

On the temporal side, the total radiation dose is divided into multiple treatment sessions (often called fractions) that span over several days [6]. Normal tissues generally recover better than tumor cells between fractions [7], making smaller doses over more fractions beneficial for sparing healthy tissue. However, prolonged treatment allows tumor proliferation, potentially necessitating higher doses for effective tumor control. To counteract this, shorter treatment schedules with fewer fractions may be preferred, depending on tumor proliferation dynamics [8]. Determining the optimal number of fractions and their timing is a critical aspect of treatment planning [9-15]. While standard protocols typically deliver one fraction per day, alternative fractionation schemes [16-21] exist, including hyperfractionation and accelerated fractionation. Equal-dose fractionation, where each fraction delivers the same dose per fraction, simplifies quality assurance. While other strategies delivering different doses in different fractions exist [22], this work adopts equal-dose fractionation for its practical advantages, assuming conventional fractionation with one fraction per day.

On the spatial side, intensity modulated radiation therapy (IMRT) [23, 24] for photon modality and intensity modulated proton therapy (IMPT) [25, 26] for proton therapy are widely used techniques to optimize dose delivery. IMRT and IMPT [27-35] allow for highly conformal dose distributions, providing better sparing of OAR. The optimization in IMRT and IMPT typically aims to minimize the discrepancy between the prescribed and actual dose delivered to the target, ensuring uniformity and adherence to clinically defined dose-volume histogram (DVH) constraints for OAR.

The problem of determining the optimal fractionation scheme, i.e., the number of fractions and dose per fraction, is a spatiotemporal optimization problem [8, 10]. When the number of fractions is fixed, optimization focuses on spatial modulation via IMRT or IMPT. However, when the number of fractions varies, biologically effective dose (BED) [36] becomes a more accurate measure of radiation's biological effect, factoring in total dose, fractionation schedule, and tumor proliferation. The linear-quadratic (LQ) model [36] is commonly used to calculate BED, that accounts for tissue-specific parameters and tumor proliferation for the target BED calculation and ignores normal tissue recovery processes for OAR BED calculation. The exact definitions used are given in Section 2.

In spatiotemporal optimization, the objective is to maximize BED to the tumor while keeping BED to OAR within acceptable limits. This optimization problem is typically quadratic and non-convex, making it computationally complex. While [10, 11, 15, 37, 38] provide closed-form solutions for cases with a single OAR, these models have limited clinical applicability since tumors are typically surrounded by multiple OAR. Moreover, most models optimize dose per fraction but do not explicitly account for spatial modulation of radiation beam intensities, limiting their practicality.

Alternative approaches [4, 12, 39] address multiple OAR and provide approximate solutions to the BED maximization problem, but these are computationally intensive. One such framework [39] generates approximate solutions for BED maximization. However, maximizing BED delivered to the target does not always yield the ideal BED for tumor control. Depending on tumor proliferation and $\alpha/\beta$ ratio, maximizing BED could result in a BED value that exceeds the ideal tumor BED or falls short, which in turn could cause excessive BED delivery to OAR or insufficient tumor control respectively.

To address this limitation, [40] propose a model that minimizes the BED delivered to OAR while ensuring that the BED delivered to the target matches the ideal BED value. This model provides a closed form solution to the problem when there is only one OAR. However, it does not explicitly model the spot intensities needed to provide a dose plan and ensure plan deliverability. This limits the practical application of the model.

Inspired by the work in [40] and to address the limitations of their model, this work proposes a model that minimizes the BED delivered to the OAR while ensuring the BED delivered to the target is close to the ideal BED value. Additionally, our model incorporates clinically relevant DVH constraints for OAR.

The model is solved multiple times for different number of fractions to determine the best choice of number of fractions from the solutions.

## 2. Problem Formulation

*2.1. Defining BED, parameters, and decision variables in our proposed optimization problem*

Parameters and decision variables:

1. $[M] = \{1, \dots, M\}$: set of indices of OAR

2. For $i \in [M], n_i$: number of voxels in OAR $i$

3. $A^i \in R^{n_i \times l}$: dose influence matrix for OAR $i$; $l$ is the number of beamlets; $A^i_j$: $j$-th row of the matrix $A^i$ that corresponds to the $j$-th voxel in OAR $i$.

4. $A^0 \in R^{n_0 \times l}$: dose influence matrix for the target; $n_0$ is the number of voxels in the target volume.

5. (Decision variable) $T$: number of fractions (days), assuming one fraction is delivered per day.

6. (Decision variable) $u \in R^k$: spot intensity vector

Biologically Effective Dose (BED) [36, 41, 22]:

1. **BED delivered to target:** Consider the function

$$\tau(T) = \frac{max\{\left((T-1) - T_l\right), 0\} \ln(2)}{T_d},$$

where $T_l$ is the time lag (in days) after which the tumor proliferation starts after treatment initiation, $T_d$ is the doubling time (in days) of the tumor. The function $\tau(T)$ accounts for the tumor proliferation/repopulation during the treatment over $T$ days. Also, for the tumor, let $\alpha_0, \beta_0$ be the parameters of the well-known LQ model used to define BED. Setting $\rho_0 = 1/(\alpha_0/\beta_0)$, under the LQ model, BED delivered to the $j$-th voxel in the tumor is defined as

$$T\left(A_j^0 u + \rho_0 \left(A_j^0 u\right)^2\right) - \tau(T).$$

2. **BED delivered to OAR:** For OAR $i$, let $\alpha_i, \beta_i$ be the parameters of the LQ-model and set $\rho_i = 1/(\alpha_i/\beta_i)$. Under the LQ model, the total BED delivered to the $j$-th voxel in OAR $i$ is

$$T\left(A_j^i u + \rho_i \left(A_j^i u\right)^2\right).$$

*2.2 Optimization problem*

The proposed optimization model is defined as

$$\begin{aligned} \min_{T,u} \sum_{i=1}^{M} w_i \frac{\sum_{j=1}^{n_i} T\left(d_j^i + \rho_i \left(d_j^i\right)^2\right)}{n_i} + f(d) \\ s.t.\ T\left(d_j^0 + \rho_0 \left(d_j^0\right)^2\right) - \tau(T) = BED_0 \quad \forall j \in [n_0] \\ d_j^i = A_j^i u \quad \forall i \in [M], j \in [n_i] \\ d_j^0 = A_j^0 u \quad \forall j \in [n_0] \\ u \in \{0\} \cup [G, +\infty\} \\ 1 \le T \le T_{max}\ and\ T\ integer. \end{aligned} \qquad (P1)$$

In problem (P1), $BED_0$ is the ideal BED value needed to ensure complete tumor cell kill for the tumor. The fourth constraint in (P1) defines a minimum-monitor-unit (MMU) value [42, 43] for $u$, where $G$ is the MMU threshold value to ensure plan deliverability. The last constraint ensures that the number of fractions is integer and less than $T_{max}$, the maximum number of fractions allowed in treatment plan. The first term in the objective function in (P1) defines the average BED value delivered to all voxels of all OAR, and the function $f(d)$ defines the least square error for the clinically used DVH constraints. The objective function, $f(d)$, is described in detail in Section 2.3. The optimization problem (P1) minimizes the BED delivered to OAR while ensuring that the BED delivered to the target is equal to the ideal BED value.

*2.3 Defining objective function $f(d)$*

The objective function, $f(d)$, is defined as

$$f(d) = \sum_{i=1}^{N_1} \frac{\mathrm{w}_1}{n_i} ||\mathrm{d}_{\Omega_{1\mathrm{i}}} - \mathrm{b}_{1\mathrm{i}}||_2^2 + \sum_{i=1}^{N_2} \frac{w_2}{n_i} ||\mathrm{d}_{\Omega_{2\mathrm{i}}} - \mathrm{b}_{2\mathrm{i}}||_2^2.$$

The function $f(d)$ in (P1) defines the least square error for the violation of the constraints defined for OAR. The value of $b$'s in the definition of $f(d)$ are the upper bounds on the acceptable total physical dose values for the OAR. These values depend on the type of constraint defined for each OAR. Each term in $f(d)$ is described below.

- The first term describes $N_1$ DVH-max constraints [44, 45] defined for OAR. For any OAR $i$, the DVH-max constraint states that at most $p$% of the total voxels in OAR $i$ should receive a dose larger than $b_{1i}$. To define this constraint, a common technique involves defining the active index set $\Omega_{1i}$ that contains indices of voxels in OAR $i$ that violate the DVH-max constraint. Mathematically, $\Omega_{1i}$ is defined as $\Omega_{1i} = \{j | j \geq p \times n_i\}$ if $d'_{p\times n_i} \geq b_{1i}$, where $d'$ is the dose distribution $d$ sorted in descending order and $n_i$ is the number of voxels in OAR $i$. Note that, $\Omega_{1i}$ is an empty set if $d'_{p\times n_i} \leq b_{1i}$. Thus, the first term in $f(d)$ defines the least square error between the actual physical dose and maximum allowed dose $b_{1i}$, for the voxels in OAR $i$ that violate the DVH-max constraint. A D-max (dose-max) constraint is defined by setting $p = 0$, ensuring that all voxels in OAR $i$ receive physical dose at most $b_{2i}$.
- The second term in $f(d)$ defines the least square error for OAR that violates the D-mean (dose-mean) constraint. For any OAR $i$, the D-mean constraint states that the mean dose delivered to all voxels in OAR $i$ should be less than or equal to $b_{2i}$. If the D-mean constraint is satisfied, the active index set $\Omega_{2i}$ is empty. However, if the constraint is not satisfied, then $\Omega_{2i} = [n_i]$. Thus, the last term defines the least square error between the actual physical dose delivered to all voxels and the maximum acceptable mean dose.

*2.4 Comparing (P1) with spatiotemporal optimization model in [39]*

The model proposed in [39] is defined as

$$
\begin{aligned}
\min_{T,u}\ & -T\left(\frac{\sum_{j=1}^{n_0} d_j^0}{n_0} + \rho_0\left(\frac{\sum_{j=1}^{n_0} d_j^0}{n_0}\right)^2\right) + \tau(T) \\
s.t.\ & T\left(d_j^i\right) + \rho_i T\left(d_j^i\right)^2 \le BED_{dv}^i \quad \forall\, i \in [N_1], j \in \Omega_{i1} \\
& T\left(d_j^i\right) + \rho_i T\left(d_j^i\right)^2 \le n_i \times BED_{mean}^i \quad \forall\, i \in [N_2], j \in \Omega_{i2} \\
& d_j^i = A_j^i u \quad \forall i \in [M], j \in [n_i] \\
& d_j^0 = A_j^0 u \quad \forall\, j \in [n_0] \\
& u \in \{0\} \cup [G, +\infty\} \\
& 1 \le T \le T_{max}\ and\ T\ integer.
\end{aligned}
\tag{P2}
$$

The optimization problem (P2) aims to maximize the BED delivered to the target while constraining the BED to the OAR within clinically defined limits. In contrast, (P1) minimizes the BED delivered to the OAR but does not explicitly enforce OAR constraints. To address this, an additional term, $f(d)$, is introduced in (P1) to approximate the satisfaction of these constraints. While (P2) ensures BED constraints for the OAR, it focuses solely on maximizing the target BED without guaranteeing that the dose is sufficient for tumor control and complete tumor cell kill. Therefore, (P1) is proposed as the more appropriate formulation, as it explicitly ensures that the BED delivered to the target meets the required threshold for tumor eradication while simultaneously optimizing the overall treatment plan.

*2.5 Solution algorithm*

To solve (P1), auxiliary variable $z$ is first added, and (P1) is redefined as

$$
\begin{aligned}
\min_{T,u}\ & \sum_{i=1}^{M} w_i \frac{\sum_{j=1}^{n_i} T\left(d_j^i + \rho_i\left(d_j^i\right)^2\right)}{n_i} + f(d) \\
s.t.\ & T\left(d_j^0 + \rho_0\left(d_j^0\right)^2\right) - \tau(T) = BED_0 \quad \forall\, j \in [n_0] \\
& d_j^i = A_j^i u \quad \forall i \in [M], j \in [n_i] \\
& d_j^0 = A_j^0 u \quad \forall\, j \in [n_0] \\
& z \in \{0\} \cup [G, +\infty\} \\
& z = u \\
& 1 \le T \le T_{max}\ and\ T\ integer.
\end{aligned}
\tag{3}
$$

Note that, Eq. (3) is a mixed integer programming problem with continuous variables $u, z$ and integer variable $T$. The problem is non-convex and computationally expensive. Since there are finite number of values that $T$ can take, $T$ is fixed to a value between 1 and $T_{max}$, and the corresponding continuous optimization problem is solved. This is done for several equally spaced values of $T$ in the range $[1, T_{max}]$, resulting in at most $T_{max}$ continuous optimization problems. Thus, Eq. (3) (with fixed value of $T$) can now be solved via iterative convex relaxation (ICR) [46, 47] and alternating direction method of multipliers (ADMM) [48, 49]. To use the ADMM method, the augmented Lagrangian of Eq. (3) is defined as

$$\begin{gathered}\min_{u,d,z} \sum_{i=1}^{M} \left( w_i \frac{\sum_{j=1}^{n_i} \; T\left(A_j^i u + \rho_i \left(A_j^i u\right)^2\right)}{n_i} + f\left(A^i u\right) \right) \\ + \frac{\mu_1}{2} ||z - u + \lambda_1||_2^2 + \frac{\mu_2}{2} ||d^0 - A^0 u + \lambda_2||_2^2 \\ s.t. \;\; T\left(d_j^0 + \rho_0 \left(d_j^0\right)^2\right) - \tau(T) = BED_0 \quad \forall\, j \in [n_0] \\ z \in \{0\} \cup [G, +\infty\}. \end{gathered} \tag{4}$$

The ICR and ADMM methods involves updating the active index sets for all terms in $f(d)$ (as described in Section 2.3), and updating each decision variable in Eq. (3) sequentially while keeping other variables fixed. Algorithm 1 describes the steps of the method and Step 4b of Algorithm 1 is described in detail in Appendix A.

**Algorithm 1: Optimization method for solving Eq. (4)**

1. **Input:** Choose parameters $\mu_1, \mu_2, w_i, w_1, w_2, w_3$
2. Initialization: Randomly initialize $u$. Choose iteration number $K$.
3. Set $\lambda_2 = d^0 = A^0 u$, $\lambda_1 = z = u$.
4. For $k = 1, \ldots, K$
    a. Find active index sets $\Omega_{1i}, \Omega_{2i}, \Omega_{3i}$ described in Section 2.3.
    b. Update primal variables $u, d, z$ one at a time by fixing all other variables and solving the resulting minimization problem.
    c. Update dual variables as follows:
    $$\lambda_1 = \lambda_1 + z - u$$
    $$\lambda_2 = \lambda_2 + d^0 - A^0 u.$$
5. **Output:** $u$

*2.6 Materials*

The proposed model (P1) and the model from literature (P2) are implemented for proton modality. The performance of (P1) is compared to (P2) for two clinical test cases: prostate and lung. Additionally, the implementation of (P1) is evaluated for three clinical scenarios: lung, prostate, and head-and-neck (HN). For IMPT implementation, the following beam angles are used: (90º, 270º), (0º, 120º, 240º), and (45º, 135º, 225º, 315º) for prostate, lung and HN cases respectively. The dose influence matrix is generated using MatRad [50] with a spot width of 5 mm on a 3 $mm^3$ dose grid. PTV-based planning is performed, adhering to clinically defined constraints for each test case.

Each test case includes two parameter settings. For the prostate case, $\alpha_0/\beta_0 = 6, 4$ Gy for the target [9, 51] and $\alpha_i/\beta_i = 3, 6$ Gy for all OARs [9, 51] with prescribed $BED_0 = 63, 69.75$ Gy. For the HN case, $\alpha_0/\beta_0 = 8, 10$ Gy for the target [52, 53] and $\alpha_i/\beta_i = 6, 3$ Gy for all OARs [12, 52] with prescribed $BED_0 = 80.92, 79.14$ Gy. For the lung case, $\alpha_0/\beta_0 = 6, 3$ Gy for the target and $\alpha_i/\beta_i = 3, 6$ Gy for all OARs and $BED_0 = 86, 106$ Gy. In all cases, the prescribed BED ($BED_0$) is based on clinically used plans: prostate (25 fractions, 1.8 Gy per fraction), HN (60 fractions, 1.09 Gy per fraction), and lung (30 fractions, 2 Gy per fraction). The $\alpha$ values are 0.2 $Gy^{-1}$ for prostate and lung cases, and 0.35 $Gy^{-1}$ for the HN case [8, 13, 52]. The doubling time ($T_d$) and lag time ($T_l$) are set to 2, 20, and 35 days, and 7, 14, and 35 days, respectively.

## 3. Results

### *3.1 Comparison of (P1) with (P2) [39]*

Figure 1 illustrates the variation in mean BED delivered to the target as a function of the number of fractions for both the proposed model (P1) and the optimal fractionation model (P2). The ideal BED required for tumor control is 63 Gy for prostate and 86 Gy for lung, based on current clinical plans. In (P1), the mean BED remains nearly constant and close to the ideal BED value, ensuring sufficient dose delivery for tumor control. In contrast, (P2) exhibits high sensitivity to the number of fractions, with BED increasing exponentially. For a small number of fractions, the mean BED falls significantly below the threshold required for complete tumor cell kill, making (P2) impractical in such cases. Additionally, in (P2), BED delivered to OAR rises rapidly with increasing fractions. While lower fraction numbers result in reduced

OAR BED, these plans are generally infeasible due to insufficient target BED. Thus, although (P2) may seem advantageous for OAR sparing, it fails to ensure adequate tumor control, making it impractical for treatment planning.

### *3.2 Evaluation of (P1) for three clinical test cases*

The performance of the proposed model (P1) is evaluated for different values of lag time ($T_l$), doubling time ($T_d$) and $\alpha/\beta$ values (as detailed in Section 2.6) to assess how the optimal number of fractions and physical dose per fraction change with tumor proliferation dynamics.

#### *3.2.1 Prostate case*

For the prostate case, two parameter sets are considered: Case 1: $(\alpha_0/\beta_0, \alpha_i/\beta_i)$ = (6,3) Gy and Case 2: $(\alpha_0/\beta_0, \alpha_i/\beta_i)$ = (4, 6) Gy. The first and second rows in Figure 2 present the model's performance for various values of $T_l, T_d$ for Case 1 and Case 2 respectively.

*Target BED stability:* Figure 2(a) shows that for all combinations of $T_l$ and $T_d$, the mean BED delivered to the target remains within 0.65% of the prescribed BED value for both cases, ensuring consistent dose delivery.

*Trend in BED delivered to OAR:* Figure 2(b) demonstrates a decreasing trend in the mean BED delivered to the bladder with an increasing number of fractions for both cases. Finally, Figure 2(c) highlights the sensitivity of the femoral head BED to fractionation. For instance, in Case 1 when $T_l = 14, T_d = 2$, the mean BED to the femoral head initially decreases with increasing fractions, reaching a minimum at 15 fractions, before rising again. A similar trend is observed for $T_l = 35, T_d = 2$, where the minimum BED is achieved at 35 fractions in Case 1.

*Optimal fractionation selection:* Considering the mean BED trends for all OARs, a suitable fractionation scheme can be determined. For example, for both Case 1 and Case 2, when $T_l = 35, T_d = 2$, choosing 35 fractions is ideal, as it minimizes the BED to both the femoral head and bladder while ensuring that the BED to the target remains close to the prescribed value. The dose per fraction determined by the output of (P1) for the prostate case when $T_l = 35, T_d = 2$ is 1.52 Gy per fraction over 35 fractions.

*3.2.2 Head-and-Neck (HN) case*

For the HN case, the following two parameter sets are considered: Case 1: $(\alpha_0/\beta_0, \alpha_i/\beta_i)$ = (6,3) Gy and Case 2: $(\alpha_0/\beta_0, \alpha_i/\beta_i)$ = (4, 6) Gy. The first and second rows in Figure 3 illustrate the model's performance for different values of $T_l, T_d$ for Case 1 and Case 2 respectively. The BED trends remain consistent across both cases.

*Target BED stability:* Figure 3(a) shows that the mean BED delivered to all target voxels stays within 0.88% of the prescribed value for every combination of $T_l$ and $T_d$ for both cases.

*Trend in BED delivered to OAR:* Figure 3(b), (c) show that, for both cases, the BED to the brainstem and brain decreases as the number of fractions increases for all values of $T_l$ and $T_d$.

*Optimal fractionation selection:* Given these observations, delivering smaller doses over a larger number of fractions is recommended for HN cases. However, to reduce patient discomfort, the maximum number of fractions is limited to 60 for all combinations of $T_l$ and $T_d$, and the output of (P1) determines the physical dose to be delivered in each fraction. For instance, when $T_l = 7, T_d = 2$, (P1) prescribes a dose of 1.31 Gy per fraction over 60 fractions for the HN case.

*3.2.3 Lung case*

For the lung case, two parameter sets are considered: Case 1: $(\alpha_0/\beta_0, \alpha_i/\beta_i)$ = (6,3) Gy and Case 2: $(\alpha_0/\beta_0, \alpha_i/\beta_i)$ = (4, 6) Gy. The first and second rows in Figure 4 present the model's performance for various values of $T_l, T_d$ for Case 1 and Case 2 respectively, showing similar trends across both cases.

*Target BED stability:* The mean BED delivered to the target remains within 0.01% of the prescribed value for all combinations of $T_l$ and $T_d$.

*Trend in BED delivered to OAR:* The BED delivered to the heart decreases with an increasing number of fractions for all parameter settings. However, the BED delivered to the lung is highly sensitive to fractionation and tumor proliferation parameters. For example, for both cases, when $T_l = 35, T_d = 2$, the mean BED to the lung initially increases with fractions up to $T = 10$. After $T = 10$, the BED rapidly decreases until $T = 35$, beyond which it stabilizes.

*Optimal fractionation selection:* Based on these observations, for the lung case, 35 fractions with a dose of 1.76 Gy per fraction, as determined by (P1), is an ideal choice for $T_l = 35, T_d = 2$, as it balances OAR dose reduction with treatment efficacy for both cases. For $T < 35$, OAR receive excessive radiation. For $T > 35$, the decrease in OAR BED is marginal, while a higher number of fractions increases patient discomfort.

### *3.3 Target dose uniformity*

Spatial optimization via $l_2$ error minimization [46] is performed on the output of (P1) to further improve dose uniformity in target and normal tissue sparing. Target dose uniformity is assessed through physical dose distributions and DVH plots for the three test cases. Figures 5–7 illustrate that incorporating spatial optimization via $l_2$ error minimization [46] enhances target dose uniformity across all cases. While spatial optimization also improves dose distribution in OARs, the improvement is marginal, indicating that the normal tissue sparing mechanism in the proposed method is already effective. The DVH plots highlight the importance of additional spatial optimization in refining dose uniformity.

#### *3.3.1 EUBED calculation*

To account for the effects of fractionation on dose distribution, the equivalent uniform biologically effective dose (EUBED) is calculated using EUBED = $-\frac{1}{\alpha} ln\left(\sum_{i=1}^{q} v_i \, e^{-\alpha BED_i}\right)$ [54], where $v_i$ is the fraction of target volume receiving $BED_i$ and $q$ is the number of dose calculation points. EUBED quantifies the impact of non-uniform dose distributions by equating them to a uniform dose that results in the same overall number of surviving cancer cells.

As shown in Figure 8, the mean BED remains within 1% of the prescribed value across all fractionation schemes, while EUBED is consistently lower than the prescribed BED. This discrepancy provides the impact of dose non-uniformity on the treatment effectiveness to ensure cell kill and suggests that a higher dose per fraction is needed to achieve the prescribed BED. Additionally, as the number of fractions increases, EUBED decreases, indicating that dose non-uniformity has a greater impact when the per-fraction dose is lower.

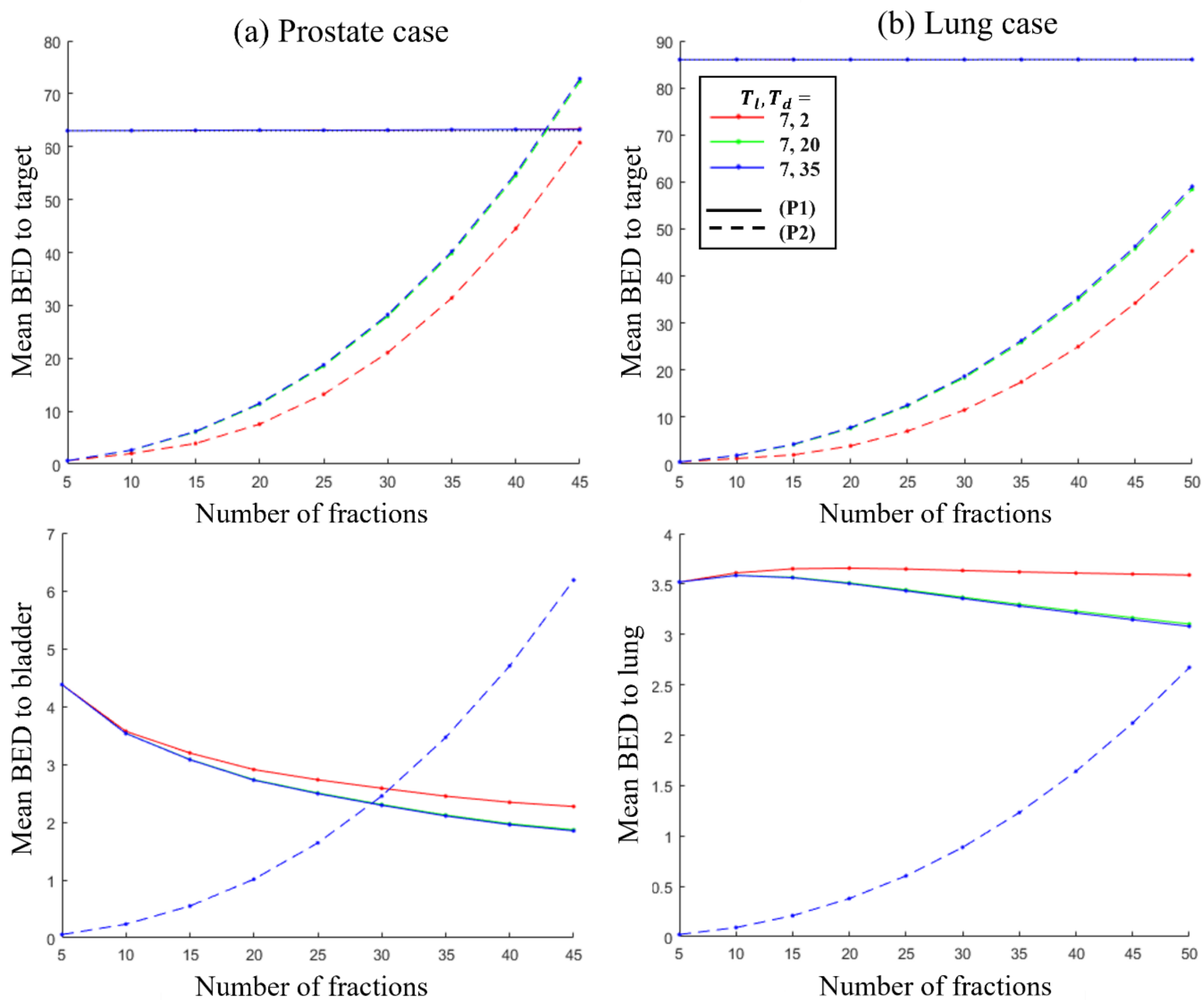

Figure 1: Plot of mean BED delivered to the target and OAR vs. number of fractions for the proposed model (P1), and optimal fractionation model (P2). First and second column show plots for prostate and lung case respectively.

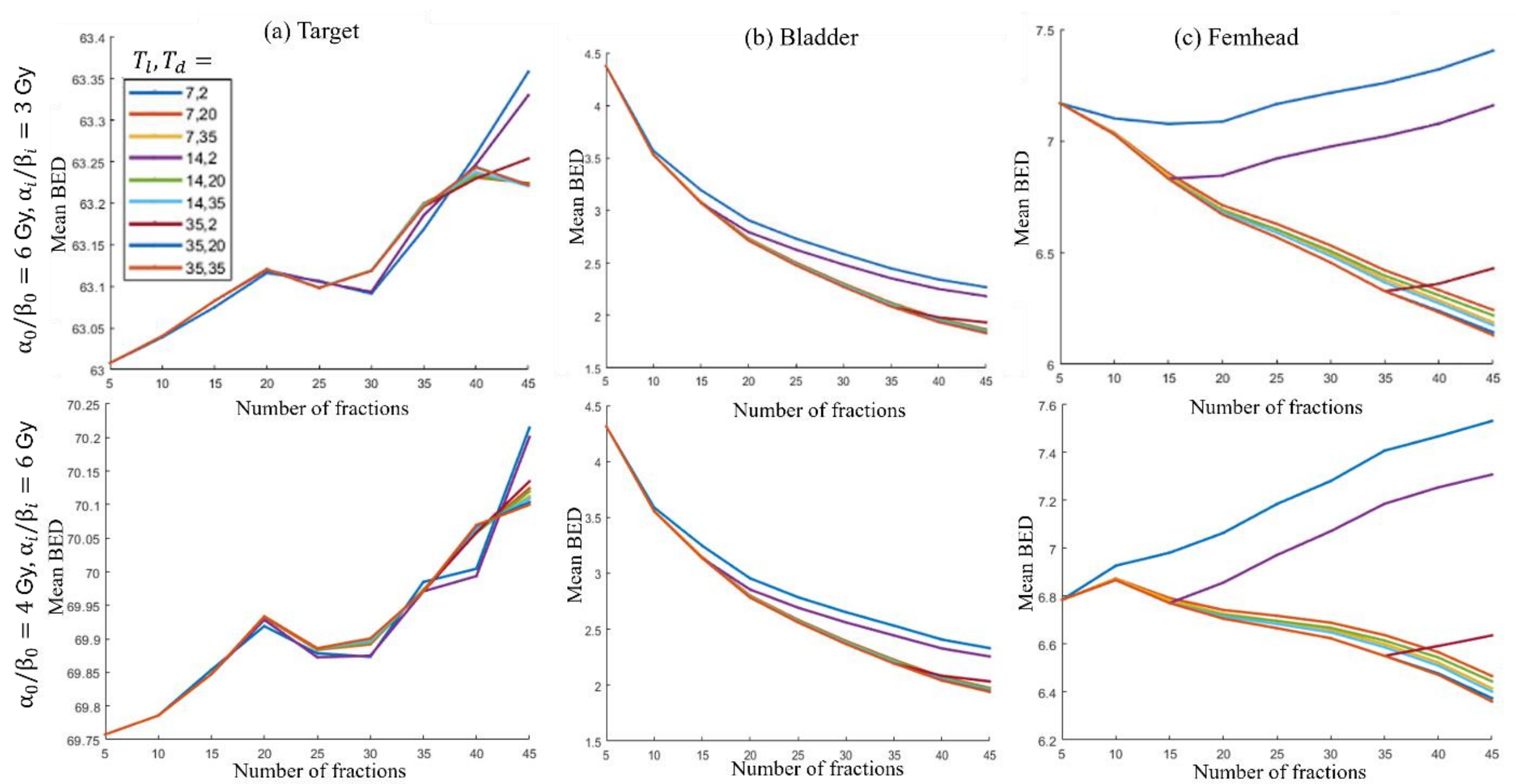

Figure 2: **Prostate.** Mean BED delivered vs number of fractions for different lag times ($T_l$) and doubling times ($T_d$) for (a) target, (b) bladder, and (c) femhead. First and second rows show results for ($\alpha_0/\beta_0, \alpha_i/\beta_i$) = (6, 3) Gy and ($\alpha_0/\beta_0, \alpha_i/\beta_i$ ) = (4, 6) Gy respectively.

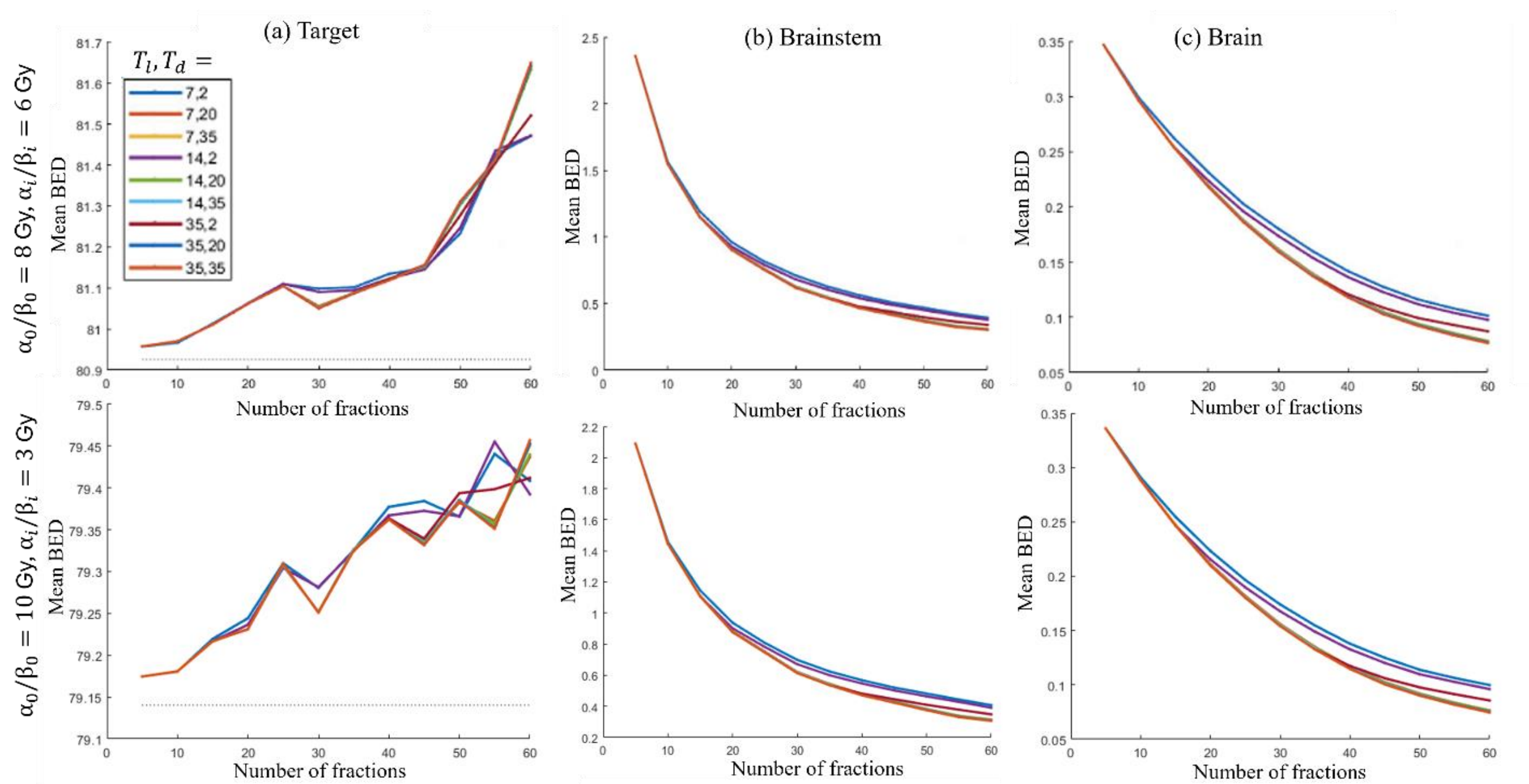

Figure 3: **HN.** Mean BED delivered vs number of fractions for different lag times ($T_l$) and doubling times (Td) for (a) target, (b) brainstem, and (c) brain. First and second rows show results for ($\alpha_0/\beta_0, \alpha_i/\beta_i$) = (8, 6) Gy and ($\alpha_0/\beta_0, \alpha_i/\beta_i$ ) = (10, 3) Gy respectively.

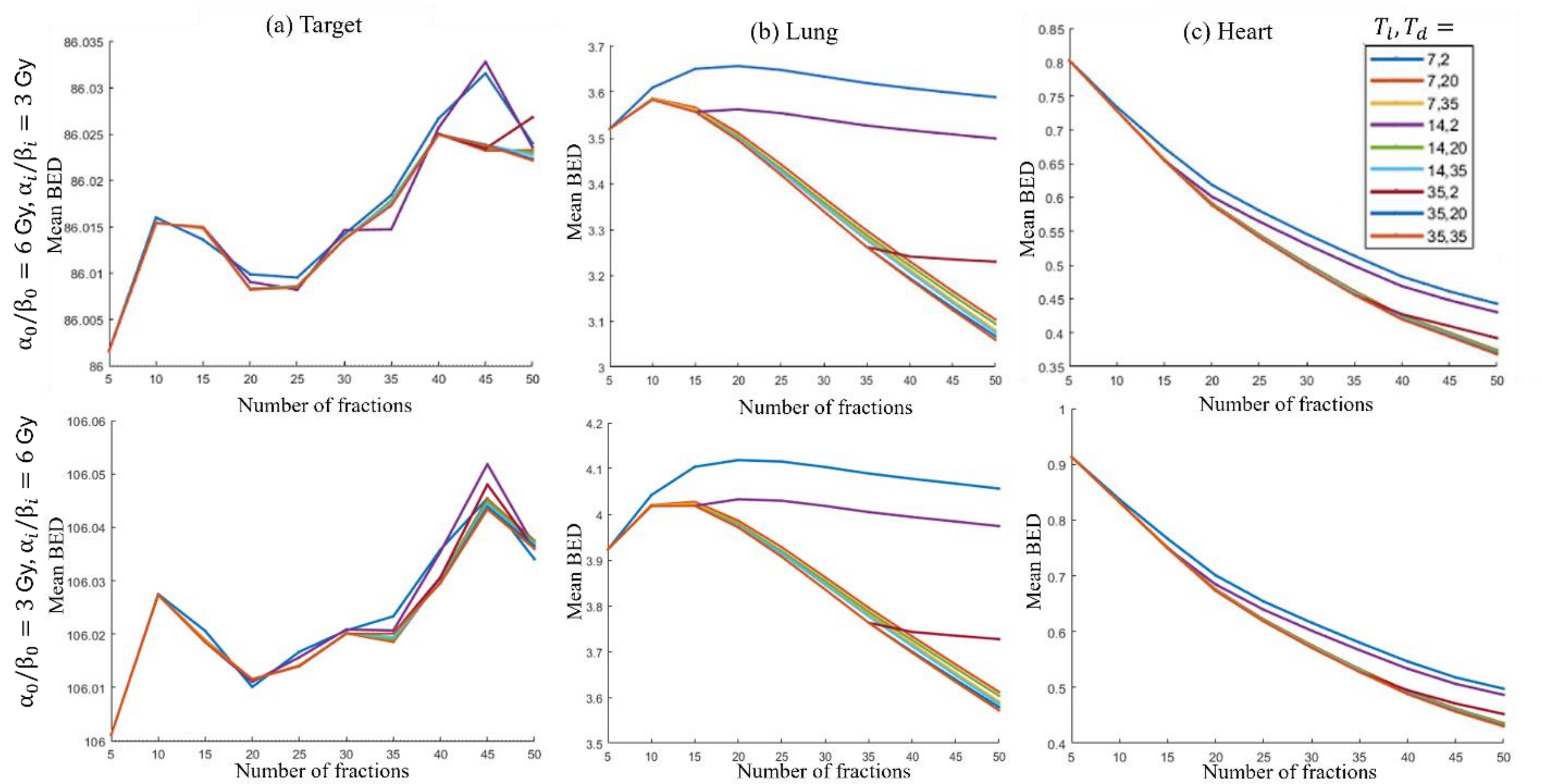

Figure 4: **Lung.** Mean BED delivered vs number of fractions for different lag times ($T_l$) and doubling times (Td) for (a) target, (b) lung, and (c) heart. First and second rows show results for $(\alpha_0/\beta_0, \alpha_i/\beta_i)$ = (6, 3) Gy and $(\alpha_0/\beta_0, \alpha_i/\beta_i\ )$ = (4, 6) Gy respectively.

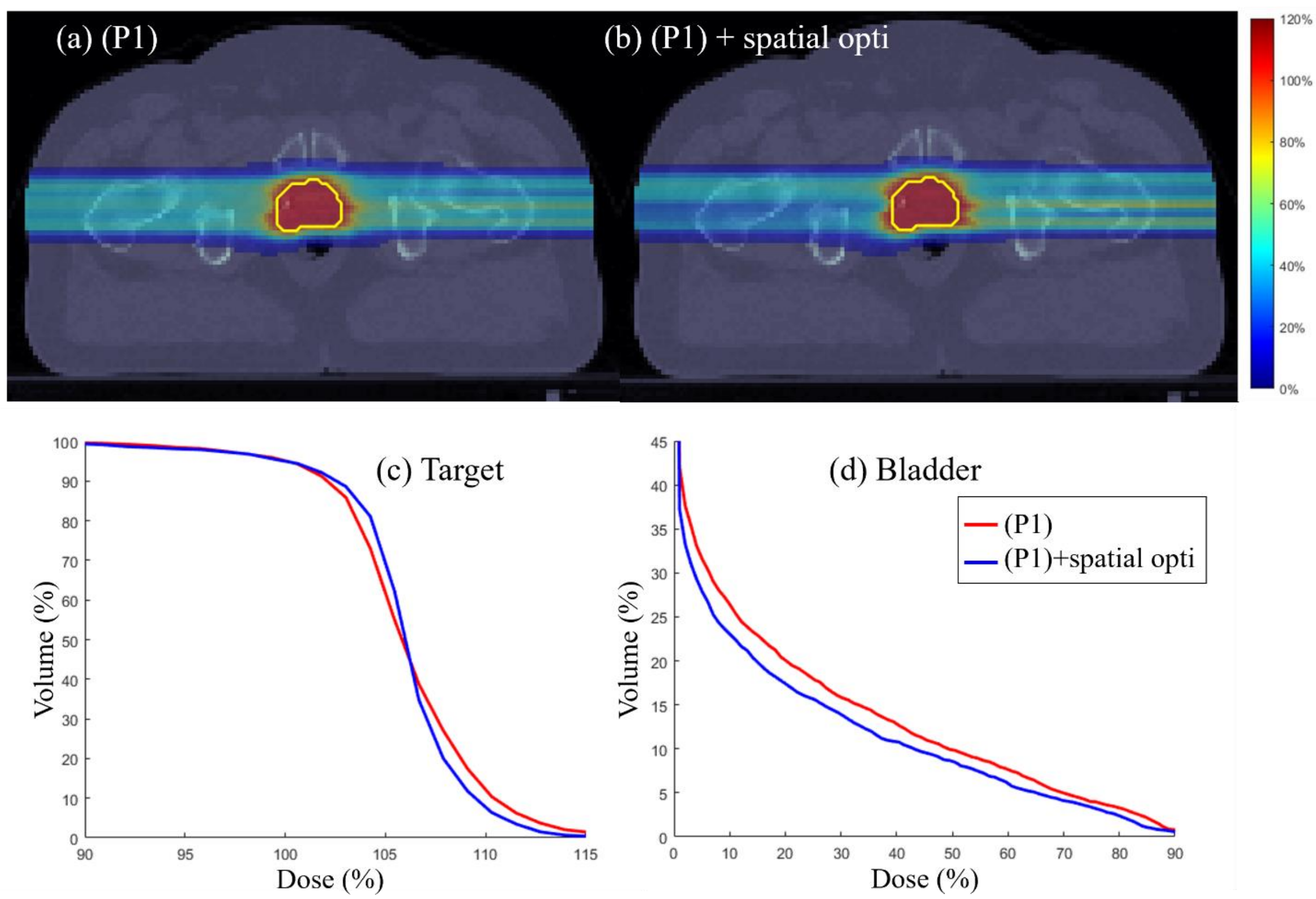

Figure 5: **Prostate.** (a) Dose plots for (P1) model, Dose plot for (P1) model followed by spatial optimization, (c) DVH plot for target, (d) DVH plot for bladder.

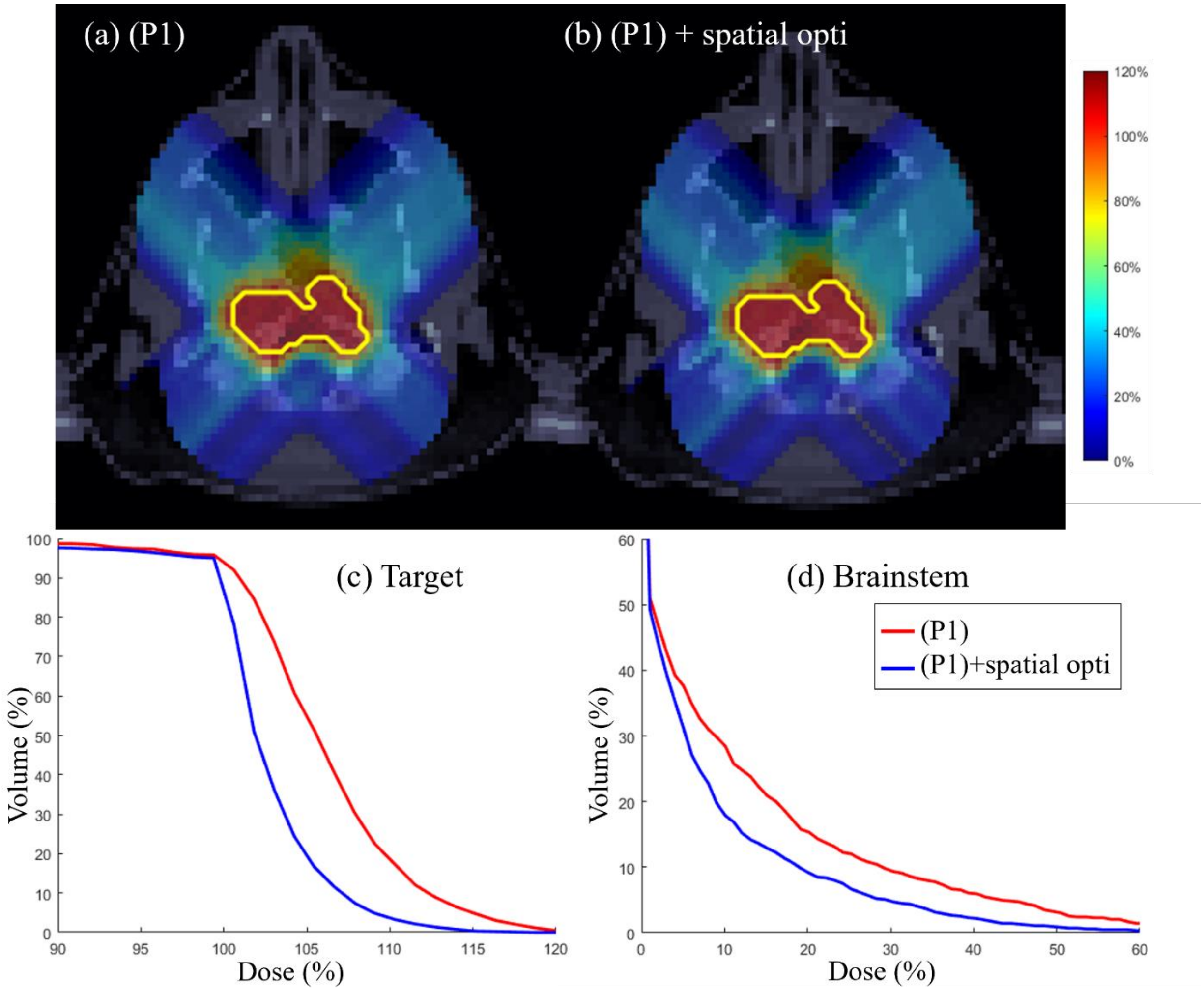

Figure 6: **HN.** (a) Dose plots for (P1) model, Dose plot for (P1) model followed by spatial optimization, (c) DVH plot for target, (d) DVH plot for brainstem.

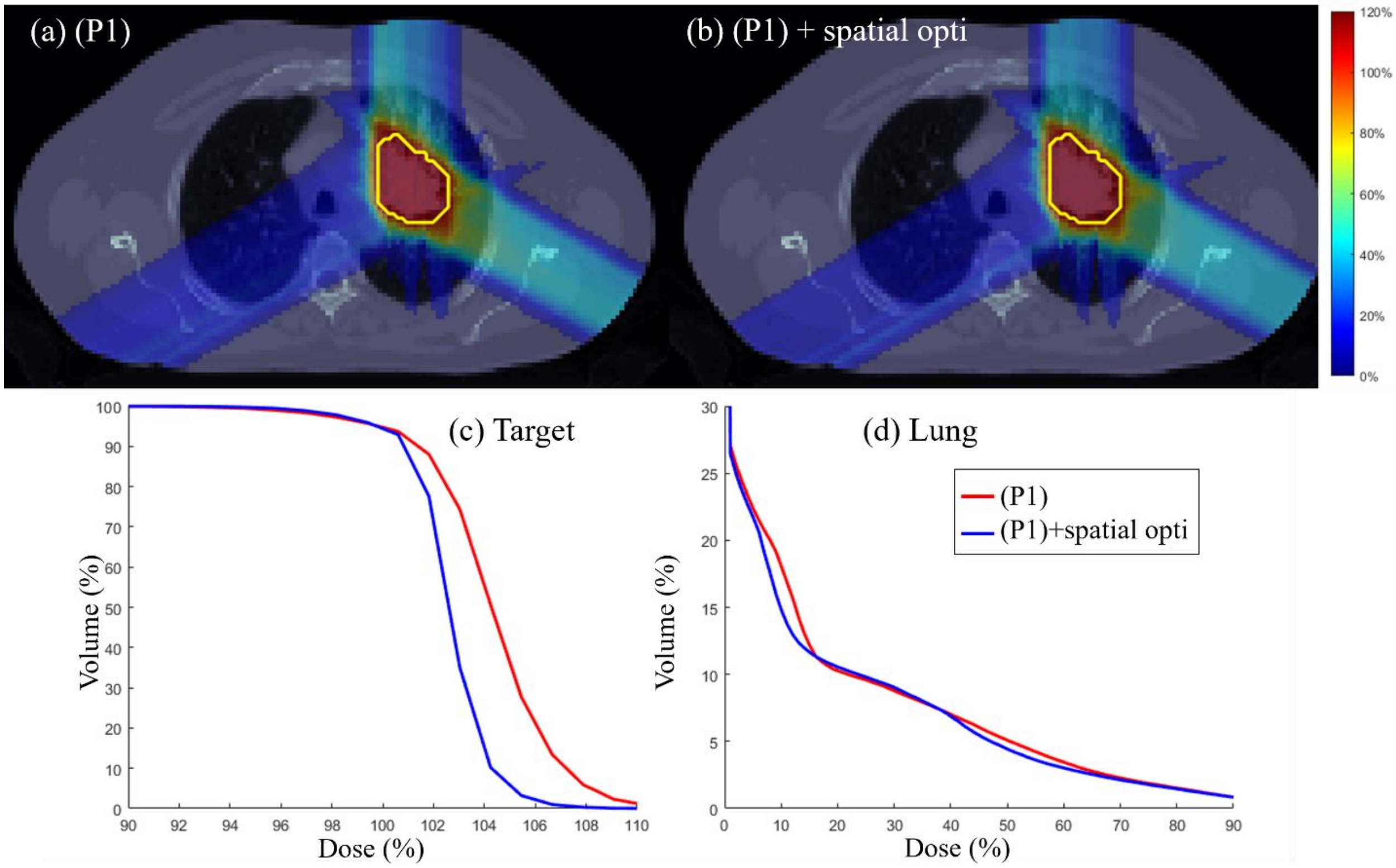

Figure 7: **Lung.** (a) Dose plots for (P1) model, Dose plot for (P1) model followed by spatial optimization, (c) DVH plot for target, (d) DVH plot for lung.

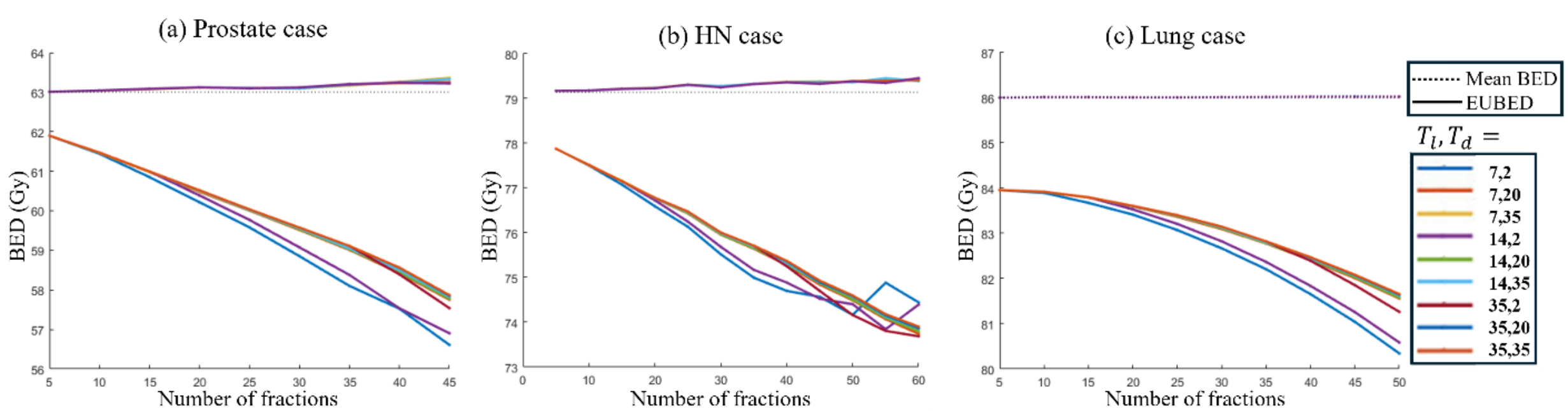

Figure 8: Comparison of mean BED and EUBED values vs. number of fractions.

## 4. Discussion

A key advantage of the proposed model over existing spatiotemporal optimization approaches is its ability to maintain a nearly constant BED delivered to the target, regardless of the number of fractions. By dynamically adjusting the dose per fraction, the model ensures that the BED remains close to the level required for complete tumor cell kill. This adaptability is crucial in preventing both insufficient and excessive BED delivery to the tumor. Additionally, the model facilitates personalized treatment planning by determining the optimal number of fractions while enhancing normal tissue sparing.

Existing models, such as (P2) [39] and the model proposed in [4], focus on maximizing the BED of the average dose delivered to all target voxels. However, this approach is less meaningful than directly optimizing or constraining the BED at the target, as done in (P1). Additionally, [4] proposes a strategy to redefine $\tau(T)$ to accommodate alternative fractionation schemes. However, we do not adopt such modifications currently, as strategies like accelerated or hyperfractionation require careful consideration of the time between fractions in the LQ model for OAR, which is often overlooked in spatiotemporal optimization.

This work assumes equal-dose fractionation to simplify quality assurance. However, nonuniform fractionation strategies that allow varying doses per fraction have been proposed [22,55,56] and shown to improve treatment outcomes. Extending our model to incorporate such strategies would increase problem size and computational complexity. Future work should assess whether the additional computational burden is justified by improved plan quality.

Finally, the current model has certain limitations. First is the dependence on the prior knowledge. The model requires accurate knowledge of the ideal BED value and tumor proliferation dynamics to generate an optimal treatment plan. If these parameters are not well-characterized, the resulting plan may not effectively balance normal tissue sparing and tumor control.

The second limitation of the current model is the slight dose non-uniformity in the target. The methodology for solving (P1) involves defining the augmented Lagrangian of (P1) (see Eq. (4)), which includes an $l_2$ error term $||d^0 - A^0 u + \lambda_2||_2^2$ for the dose delivered to the target voxels. Thus, solving Eq. (4) via Algorithm 1 ensures that the physical dose $d^0$ per fraction delivered over $T$ fractions achieves the ideal BED for the target, while the $l_2$ error term ensures that the corresponding spot intensities $u$ deliver a dose *nearly* equal to $d^0$. Thus, the solution to Eq. (4) introduces some dose non-uniformity. Additional spatial dose optimization using classical optimization models in IMPT or IMRT needs to be performed to improve overall plan quality. Finally, there is a need for further validation. More rigorous testing is required to fully establish the efficacy of the model in clinical settings.

The third limitation is the computational complexity of solving (P1). An exact solution would require optimizing both the number of fractions and spot intensities simultaneously, but the model's high nonconvexity makes this infeasible. Instead, the proposed methodology solves multiple instances of the problem, each with a fixed number of fractions. As a result, the computational complexity increases with the number of possible fractionation choices, limiting the practical applicability of the current solution methodology.

Despite these limitations, the current results validate the effectiveness of clinically used dose plans and provide insights into how fractionation impacts normal tissue sparing. Thus, this model serves as a foundational framework for spatiotemporal optimization, offering a structured approach to personalized radiation therapy planning.

## 5. Conclusion

This work presents a spatiotemporal optimization model (P1) designed to minimize the BED delivered to OAR while ensuring the target receives a BED value close to the ideal value necessary for complete tumor cell kill. The model outputs an optimal treatment plan, including the number of fractions and the physical dose per fraction. Results demonstrate that the proposed approach effectively determines the optimal number of fractions across three clinical test cases, accounting for varying tumor proliferation dynamics.

## Appendix A: Step 4b of Algorithm 1 for solving the Augmented Lagrangian formulation (Eq. (4))

In this section, Step 4b of Algorithm 1 is described in detail. At each iteration $k$ of the algorithm, the primal variables are updated as described below.

1. Updating $u$: At each iteration, fix all variables except $u$ in Eq. (4). The resulting minimization problem is unconstrained in $u$. We can then take first-order derivative of the objective function with respect to $u$ and set it to 0. The solution to the resulting linear system of equations provides the value of $u$.

2. Updating $z$: At each iteration, fix all variables except $z$ in Eq. (4). The problem has a closed form solution defined using soft thresholding as follows:

$$z = \begin{cases} max\,(G, u - \lambda_1), & if\ u - \lambda_1 \geq G/2 \\ 0, & otherwise. \end{cases}$$

3. Updating $d^0$: Fix all variables except $d^0$ in Eq. (4). The resulting problem is

$$\min_{u,d,z} ||d^0 - A^0 u + \lambda_2||_2^2$$
$$s.t.\ \ T\left(d_j^0 + \rho_0\left(d_j^0\right)^2\right) - \tau(T) = BED_0 \quad \forall\, j \in [n_0] \tag{5}$$

The equality constraint in Eq. (5) can be equivalently written as $\left(d_j^0 + \frac{1}{2\rho_0}\right)^2 = \frac{BED_0 + \tau(T)}{T\rho_0} + \frac{1}{4\rho_0^2}$. The optimal solution to Eq. (5) is the projection of $A^0 u - \lambda_2$ on the equality constraint. The projection is defined as $d^0 = (1+q)(A^0 u - \lambda_2) + q\frac{1}{2\rho_0}$, where $q = \sqrt{\frac{\frac{BED_0+\tau(T)}{T\rho_0} + \frac{1}{4\rho_0^2}}{\left((A^0 u - \lambda_2 + \frac{1}{2\rho_0}\right)^2}} - 1$.